\documentstyle[stwol,epsfig]{article}

\input{psfig}

\def\Journal#1#2#3#4{{#1} {\bf #2}, #3 (#4)}

\def\NPB{{\em Nucl. Phys.} B}
\def\PLB{{\em Phys. Lett.}  B}

\def\PRD{{\em Phys. Rev.} D}

\def\be{\begin{equation}}
\def\ee{\end{equation}}
\def\bea{\begin{eqnarray}}
\def\eea{\end{eqnarray}}

\begin{document}

\title{DYNAMICAL DETERMINATION OF THE TOP QUARK AND HIGGS
MASSES AND FERMION MASSES IN THE
ANTI-GRAND UNIFICATION MODEL}

\author{  C.D. FROGGATT }

\address{ Department of Physics and Astronomy,
 Glasgow University, Glasgow G12 8QQ, Scotland}

\author{ H.B. NIELSEN, D.J. SMITH }

\address{ The  Niels  Bohr Institute,
Blegdamsvej 17, DK-2100 Copenhagen {\O}, Denmark}


\twocolumn[\maketitle\abstracts{
The multiple point criticality principle is
applied to the pure Standard Model (SM),
with a desert up to the Planck scale. We are
thereby led to impose the
constraint that the effective Higgs potential
should have two degenerate minima, one of which
should have a vacuum expectation value of
order unity in Planck units. In this way we
predict a top quark mass of $173 \pm 5$ GeV
and a Higgs particle mass of $135 \pm 9$ GeV.
The quark and lepton mass matrices are considered
in the anti-grand unified extension of the SM
based on the gauge group $SMG^3 \otimes U(1)_f$;
this group contains three copies of the SM
gauge group SMG, one for each generation,
and an abelian flavour group $U(1)_f$.
The 9 quark and lepton masses and 3
mixing angles are fitted
using 3 free parameters, with the overall mass scale
set by the electroweak interaction.
It is pointed out that the same results can be obtained in an
anomaly free $SMG \otimes U(1)^3$ model.}]

\section{Introduction}
\label{sec:intro}

On rather general grounds, we expect that all
coupling constants become dynamical in quantum gravity
due to the non-local effects of baby universes.
\cite{baby}.
We have argued \cite{corfuhbn} that
the mild form of non-locality in quantum gravity,
respecting reparameterisation invariance \cite{book},
leads to the realisation in Nature of the ``multiple
point criticality principle''. According to this
principle, Nature should choose coupling constant
values such that the vacuum can exist in
degenerate phases, in a very similar way to the stable
coexistence of ice, water and vapour (in a thermos
flask for example) in a mixture with fixed energy and number of molecules.
We apply this principle to the pure Standard Model (SM) with
a desert up to the Planck scale, and
give a precise dynamical determination of the top quark mass $M_t$
and the Higgs particle mass $M_H$.

The top quark is the only SM fermion which has a
mass of order the electroweak scale and a SM Yukawa
coupling of order unity. All of the other fermion masses
are suppressed relative to the natural scale of the SM.
The Yukawa couplings range over five orders of magnitude,
from of order $10^5$ for the electron to of order unity
for the top quark. It is this fermion mass hierarchy
problem that we address in the second half of the talk.
We introduce an extension of
the SM where such a large range of
Yukawa couplings is natural, due to the existence of new
approximately conserved chiral gauge quantum numbers
which protect the fermions from gaining a mass.
This so-called
anti-grand unification model \cite{Bennett} is
based on the gauge group $SMG^3 \otimes U(1)_f$, where
$SMG=SU(3) \otimes SU(2) \otimes U(1)$.
In our model all the elements of the Yukawa
matrices (except for one element which leads
to the unsuppressed top mass) are
suppressed (relative to the assumed natural order 1)
by a product of the
ratios of the vacuum expectation values (VEVs) of the
Higgs fields required to break the extended gauge symmetry
to the fundamental (Planck) scale
of the theory. In this way we can
express very small numbers such as $10^{-5}$
as the product of 5 numbers of
order $10^{-1}$.

\section{The Ice-Water Analogue}
\label{sec:ice}

In the analogy of the ice, water and vapour system,
the important point for us is
that by enforcing fixed values of the extensive quantities, such as
energy, the number of moles and the volume, you can very likely
come to make such a choice of these values that a mixture
has to occur. In that case then the temperature and pressure (i.~e.~the
intensive quantities) take very specific values, namely the values
at the triple point. We want to stress that this phenomenon of
thus getting specific intensive quantities only happens for
first order phase transitions, and it is only {\em likely} to
happen for rather strongly first order phase transitions.
By strongly first order, we here mean that the interval of values
for the extensive
quantities which do not allow the existence of
a single phase is rather large.
Because the phase transition between water and ice is first order,
one very often finds slush (partially melted snow or ice) in winter
at just zero degree celsius. And conversely
you may guess with justification that if the temperature happens
to be suspiciously close to
zero, it is because of the existence of such a mixture: slush.
But for a very weakly first
order or second order phase transition, the connection with
a mixture is not so likely.

In the analogy considered in this paper the coupling constants,
such as the
Higgs self coupling and the top quark Yukawa coupling,
correspond to intensive
quantities like temperature and pressure.
Now it is well-known that the pure Standard Model, with
one loop corrections say, can have two minima in the
effective Higgs field potential. If really there were
some reason for Nature to require phase coexistence,
it would be expected that the ``vacua'' corresponding to these
minima should be able to energetically coexist, which
means that they should be degenerate.
If the vacuum degeneracy
requirement should have a good chance of being relevant, the ``phase
transition'' between the two vacua should be strongly first order.
That is to say there should be an appreciable interval of extensive variable
values leading to a necessity for the presence of the two phases in the
Universe. Such an extensive variable might be
e.~g.~$\int d^4x |\phi(x)|^2 $.
If, as we shall assume, Planck units reflect the fundamental
physics, it would be natural to interpret this strongly first order
transition requirement to mean that, in Planck units, the extensive
variable densities
$\frac{\int d^4x |\phi(x)|^2}{  \int d^4x }$ = $<|\phi|^2>$
for the two vacua should differ by a quantity of order unity.
Phenomenologically $|\phi|^2$ is very small in Planck
units for the vacuum in which we live, and hence the
other vacuum should have
$|\phi|^2$ of order unity in Planck units.

\begin{figure}[h]
\leavevmode
\centerline{
\psfig{file=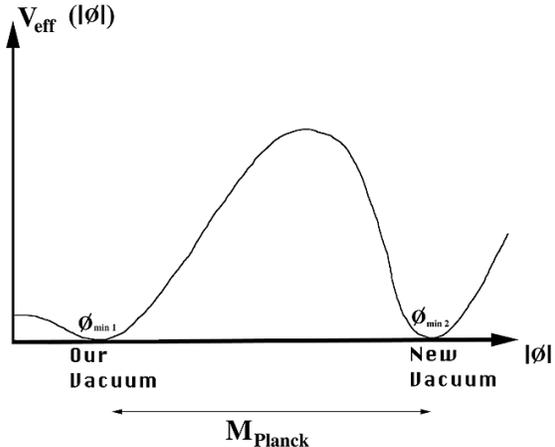,width=8.5cm,%
bbllx=0pt,bblly=0pt,bburx=548pt,bbury=444pt,%
clip=}
}
\caption{
This symbolic graph of the effective potential
$V_{\rm{eff}}(\phi)$ for the Standard Model Higgs field
illustrates the two assumptions which lead to our prediction of
the top quark and Higgs boson masses:
1) Two equally deep minima, 2) achieved for $|\phi|$
values differing, order of magnitudewise, by
unity in Planck units.
}
\label{fig:veff}
\end{figure}

Thus the application of the multiple point criticality principle
to the pure SM
leads to our two crucial assumptions:
\newline
a) The two minima in the SM effective Higgs potential
are degenerate:
$V_{\rm{eff}}(\phi_{\rm{min}\; 1}) = V_{\rm{eff}}(\phi_{\rm{min} \; 2})$.
\newline
b) The second minimum has a
Higgs field squared:
$<|\phi_{\rm{min} \; 2}|^2> = {\cal O}(M_{\rm{Planck}}^2)
\sim (10^{19}$ GeV)$^2$.
\newline
These assumptions are illustrated in Figure 1 and in the
next section we show how they lead to precise predictions for
$M_t$ and $M_H$.

\section{Calculation of the Higgs and Top Masses}
\label{sec:calc}

We use the renormalisation group improved
effective potential,
identifying
the renormalisation point with the field strength $\phi$:
\begin{equation}
V_{\rm{eff}}(\phi) \; = \; \frac{1}{2}m_{h}^2(\mu = |\phi |)\,
|\phi |^2 \; + \; \frac{1}{8}\lambda (\mu = |\phi | )\, |\phi |^4
\end{equation}
Now the condition
$V_{\rm{eff}}(\phi_{\rm{min}\; 1})  = V_{\rm{eff}}(\phi_{\rm{min} \; 2})$,
where one of the minima corresponds to our vacuum with
$\phi_{\rm{min}\; 1} = 246$ GeV,
defines (part of) the vacuum stability curve in the $M_t-M_H$ plane,
for $\phi_{\rm{min} \; 2} < \Lambda$. Here $\Lambda$ is the physical
cut-off scale for the pure SM beyond which new physics enters.
We take $\Lambda \simeq M_{\rm{Planck}} \sim 10^{19}$ GeV
and the vacuum stability curve for this case is illustrated
\cite{casas} in Figure 2
\begin{figure}
\leavevmode
\vspace{-0.5cm}
\centerline{\epsfig{file=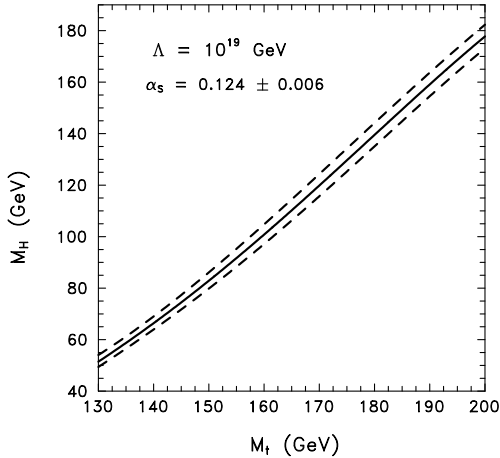,width=7cm,%
bbllx=95pt,bblly=130pt,bburx=510pt,bbury=555pt,%
angle=270,clip=}
}
\caption{SM vacuum stability curve
for $\Lambda = 10^{19}$ GeV and $\alpha_s = 0.124$ (solid line),
$\alpha_s = 0.118$ (upper dashed line), $\alpha_s = 0.130$
(lower dashed line).}
\label{fig:vacstab}
\end{figure}

If we now also impose the strong first order transition requirement,
$\phi_{\rm{min}\; 2} = {\cal O}(\rm{M}_{\rm{Planck}})$
we are interested in the
situation when $\phi_{\rm{min} \; 2}\gg\phi_{\rm{min}\; 1}$. In
this case the energy density in our vacuum 1 is exceedingly small
compared to $\phi^4_{\rm{min} \; 2}$. Also the $|\phi|^4$ term
will a priori strongly dominate the $|\phi|^2$ term
in vacuum 2.
So we basically get the vacuum
degeneracy condition to mean that, at the
vacuum 2 minimum, the effective coefficient $\lambda(\phi_{\rm{min} \; 2})$
must be zero with high accuracy.  At the same $\phi$-value the derivative
of the effective potential $V_{\rm{eff}}(\phi)$ should be zero, because
it has a minimum there. In the approximation $V_{\rm{eff}}(\phi) \approx
\frac{1}{8}\lambda(\phi) \phi^4 $ the derivative of $V_{\rm{eff}}(\phi)$
with respect to $\phi$ becomes
\begin{equation}
\frac{dV_{\rm{eff}}(\phi)}{d\phi}|_{\phi_{\rm{min} \; 2}}
= \frac{1}{2}\lambda(\phi)\phi^3
+\frac{1}{8}\frac{d\lambda(\phi)}{d\phi}\phi^4
=\frac{1}{8}\beta_{\lambda} \phi^3
\end{equation}
and thus at the second minimum the beta-function
$\beta_{\lambda}$
vanishes, as well as $\lambda(\phi)$.
So we imposed the conditions $\beta_{\lambda}=\lambda=0$
near the Planck scale,
$\phi_{\rm{min} \; 2} \simeq M_{\rm{Planck}}$.
Using the two loop renormalisaion group equations, these
two conditions determine a single point on the vacuum
stability curve, giving our Standard Model
criticality prediction for both the top quark and
Higgs boson pole masses \cite{smtop}:
\begin{equation}
M_{t} = 173 \pm 5\ \mbox{GeV} \quad M_{H} = 135 \pm 9\ \mbox{GeV}.
\end{equation}

\section{The Anti-grand Unification Model}
\label{SMG3U1}
The anti-grand unified gauge group is:
\begin{equation}
G=SMG_1 \otimes SMG_2 \otimes SMG_3 \otimes U(1)_f
\end{equation}
\begin{equation}
\mbox{where} \quad SMG_i=SU(3)_i \otimes SU(2)_i \otimes U(1)_i
\end{equation}
The three $SMG_i$ groups are broken down to their diagonal
subgroup---the usual SM gauge group SMG---and
the $U(1)_f$ group is totally broken near the Planck scale.
As discussed by Holger Nielsen at this conference,
the multiple point criticality principle applied to this
model successfully predicts the values of the
three running SM gauge coupling constants \cite{Bennett}.

We put the SM fermions into this group in an obvious way. We have one
generation of fermions coupling to each $SMG_i$
in exactly the same way as they
would couple to the SMG in the SM.
We then choose $U(1)_f$ charges with the constraint that there should
be no anomalies and no new mass-protected fermions, giving
the essentially unique
set of charges shown in Table~\ref{Q_f}. We have labelled the fermions
coupling to $SMG_i$ by the names of the `i'th
proto-generation of SM fermions.

\begin{table}[h]
\caption{$U(1)_f$ charges of the fermions.}
\begin{displaymath}
\begin{array}{|c|c|c|c|c|c|}
\hline
{\rm Fermion} & Q_f & {\rm Fermion} & Q_f & {\rm Fermion} & Q_f \\ \hline
u_L & 0 & c_L & 0 & t_L & 0 \\ \hline
u_R & 0 & c_R & 1 & t_R & -1 \\ \hline
d_R & 0 & s_R & -1 & b_R & 1 \\ \hline
e_L & 0 & \mu_L & 0 & \tau_L & 0 \\ \hline
e_R & 0 & \mu_R & -1 & \tau_R & 1 \\ \hline
\end{array}
\end{displaymath}
\label{Q_f}
\end{table}
Now we must choose appropriate Higgs fields to break $G$ down to the SMG.
The quantum numbers of the fermion fields are determined by the
theoretical structure of the model (in particular the requirement of
anomaly cancellation), but we do have some freedom in the choice
of the quantum numbers of the Higgs fields.
We choose the Higgs fields so as to get a realistic model of
the suppression of fermion masses. In fact we specify the U(1)
charges of the Higgs fields and use a natural generalisation
of the usual SM charge quantisation rule to determine the
non-abelian representations.
The Weinberg-Salam Higgs field $\phi_{WS}$ takes
highly non-trivial quantum numbers under $G$.
In this way we determine
an order of magnitude representation of the Yukawa coupling
matrices, in terms of three Higgs field VEVs; W, T and $\xi$
measured in Planck units. Since we only use the
abelian charges, the same results can be obtained in an anomaly free
$SMG\otimes U(1)^3$ model.

\section{Mass Matrices}
\label{matrices}
The fermion mass matrices are expressed in terms of Yukawa
matrices and $<\phi_{WS}> = 246$ GeV
by:
$M_f = Y_f <\phi_{WS}>/\sqrt{2}$.
Our model \cite{smg3m} gives the following order of magnitude Yukawa
matrices at the Planck scale,
where unknown complex coefficients of $\cal O$(1)
in the matrix elements have been ignored.
\begin{eqnarray}
\label{M_u}
Y_u & \simeq & \left(\begin{array}{ccc} WT^2\xi^2 & WT^2\xi & W^2T\xi \\
				   WT^2\xi^3 & WT^2    & W^2T    \\
				   \xi^3     & 1       & WT \end{array}
	  \right) \\
\label{M_d}
Y_d & \simeq & \left(\begin{array}{ccc} WT^2\xi^2 & WT^2\xi & T^3\xi  \\
				   WT^2\xi   & WT^2    & T^3     \\
				   W^2T^4\xi & W^2T^4  & WT \end{array}
	  \right) \\
\label{M_l}
Y_l & \simeq & \left(\begin{array}{ccc} WT^2\xi^2 & WT^2\xi^3 & WT^4\xi \\
				   WT^2\xi^5 & WT^2    & WT^4\xi^2 \\
				   WT^5\xi^3 & W^2T^4  & WT \end{array}
	  \right)
\end{eqnarray}
We note that these mass matrices are not symmetric but they
do have some important features. The diagonal elements in all
3 mass matrices are the same, giving the SU(5)-like results
$m_b\approx m_{\tau}$ and $m_s \approx m_{\mu}$, but the
off-diagonal elements dominate $Y_u$ making $m_t$ and $m_c$
respectively larger. The diagonal and off-diagonal contributions
to the lowest eigenvalue of $Y_d$ are approximately equal, giving
$m_d \ge m_u \approx m_e$.
A three parameter order of magnitude fit with
W = 0.158, T = 0.081 and $\xi$ = 0.099
successfully reproduces the observed masses and mixing angles
within a factor of two---see Table 2.
\begin{table}[h]
\caption{Best fit to experimental data. All masses are running masses at 1 GeV
except the top quark mass which is the pole mass.}
\begin{displaymath}
\begin{array}{|c|c|c|}
\hline
 & {\rm Fitted} & {\rm Experimental} \\ \hline
m_u & 3.8 {\rm \; MeV} & 4 {\rm \; MeV} \\ \hline
m_d & 7.4 {\rm \; MeV} & 9 {\rm \; MeV} \\ \hline
m_e & 1.0 {\rm \; MeV} & 0.5 {\rm \; MeV} \\ \hline
m_c & 0.83 {\rm \; GeV} & 1.4 {\rm \; GeV} \\ \hline
m_s & 415 {\rm \; MeV} & 200 {\rm \; MeV} \\ \hline
m_{\mu} & 103 {\rm \; MeV} & 105 {\rm \; MeV} \\ \hline
M_t & 187 {\rm \; GeV} & 180 {\rm \; GeV} \\ \hline
m_b & 7.6 {\rm \; GeV} & 6.3 {\rm \; GeV} \\ \hline
m_{\tau} & 1.32 {\rm \; GeV} & 1.78 {\rm \; GeV} \\ \hline
V_{us} & 0.18 & 0.22 \\ \hline
V_{cb} & 0.029 & 0.041 \\ \hline
V_{ub} & 0.0030 & 0.002 - 0.005 \\ \hline
\end{array}
\end{displaymath}
\label{bestfit}
\end{table}

\section{Conclusion}
Our determination of the pole masses,
\linebreak
$(M_t,M_H) = (173 \pm 5, 135 \pm 9)$ GeV
follows from 3 assumptions:
\newline
1. The pure SM is valid, without any supersymmetry, up to
the Planck scale.
\newline
2. The SM Higgs potential $V_{\rm{eff}}$ has 2 degenerate minima.
\newline
3. There is a strong first order phase transition, requiring
$\phi_{\rm{min} \; 2} \simeq M_{\rm{Planck}}$.
\newline
The coexistence of these 2 vacuum phases in
spacetime suggests a mild breakdown of locality
as in baby universe theory.

The anti-grand unified $SMG^3 \otimes U(1)_f$ model, or the
effective $SMG \otimes U(1)^3$ model, successfully fits
9 fermion mases and 3 mixing angles in terms of 3 Higgs VEVs.

\end{document}